\documentclass[11pt]{article}
\usepackage[margin=1in]{geometry}
\usepackage{graphicx}
\usepackage{threeparttable}
\usepackage{cite}
\usepackage{amsmath}
\usepackage{amsmath,amsfonts,amssymb}
\usepackage{flexisym}
\usepackage{mathtools}
\usepackage{xspace}
\usepackage{xfrac}
\usepackage[onehalfspacing]{setspace}

\title{Patient Diversion Across Primary Health Centers Using Real Time Delay Predictors}
\author{Najiya Fatma$^{1}$, Varun Ramamohan$^{1}$  \\
        \small $^{1}$Department of Mechanical Engineering, Indian Institute of Technology Delhi, Hauz Khas, New Delhi \\
}
\date{January 2021}

\begin{document}

\maketitle
\begin{abstract}
In the current work, we consider diversion of childbirth patients who arrive seeking emergency admission to public primary health centers (PHCs). PHCs are the first point of contact for an Indian patient with formal medical care, and offer medical care on an outpatient basis, and limited inpatient and childbirth care. In this context, real-time prediction of the wait time of the arriving patient becomes important in order to determine whether the patient must be diverted to another PHC or not. We study this problem using a discrete event simulation that we develop of medical care operations in two PHCs in India. We approximate the labour room service at each PHC as an M/G/1 queueing system and show how the accuracy of real-time delay predictors impacts the extent of the change in operational outcomes at each PHC. We simulate patient diversion using actual delays as well as the delay estimates generated by various delay predictors based on the state of the system such as queue-length, elapsed service time, and observed delay histories. The simulation of the diversion process also incorporates travel time between the PHCs. We also propose a new delay predictor that incorporates information regarding the system state as well as the service time distribution. We compare the operational outcomes at both PHCs without diversion and with diversion using the above delay predictors. We show numerically that more accurate delay predictors lead to more equitable distribution of resources involved in provision of childbirth care across both PHCs.
\end{abstract}

\section{Introduction}
The number of patients accessing healthcare services has increased significantly across the world, including in India \cite{poon2018trends, shrivastava2017obesity}, and subsequently demand at all tiers of healthcare facilities is likely to grow in the coming years. Long wait times burden the healthcare administration in addition to inconveniencing and worsening patient outcomes, and to alleviate this, referral mechanisms of various types are implemented. Inter-facility referral systems are considered to be effective mechanisms for reducing delays in admission to hospitals \cite{pouramin2020delays}, and strengthening the efficiency of referral system has potential to improve quality of care in the community \cite{give2019strengthening}. Two types of referral policies \cite{handayani2018health} are typically practiced in the healthcare context: (a) vertical referral, when the required equipment and/or expertise are unavailable at the current healthcare facility, and therefore patients are referred to a higher level of care, and (b) horizontal referral, when patients cannot access treatment within some threshold time duration due to limited healthcare capacity and are therefore referred to other facilities typically at a similar level of care.  In this work, we consider horizontal referral, and we use the term patient diversion in place of horizontal referral to be consistent with the health operations literature. We study patient diversion in the context of primary health center (PHC) operations in the Indian context. This is because previous studies \cite{shoaib20} have shown that a significant proportion of childbirth patients are likely to not receive care at public primary healthcare facilities in the Indian context within a reasonable timeframe (e.g., two hours). Hence, we investigate whether diverting these patients to other primary healthcare facilities can reduce their estimated wait times.
\noindent
\\
\\
While patient diversion has been proposed in many studies as a method to reduce wait times, particularly in the context of ambulance diversion to reduce emergency department delays 
\cite{yarmohammadian2017overcrowding}, we propose real-time delay prediction as a basis for making the diversion decision. This is because estimating the delay for a given patient arriving at the facility seeking care as a function of the state of the system or delay histories on a real-time basis (i.e., at the time the patient arrives at the facility) can provide the most up to date information that can inform the diversion decision, as opposed to using steady state measures of average wait time. Note that the delay prediction must be made at all facilities to which the patient is being considered for diversion, and not only at the facility they first arrive at. In this work, we estimate real-time delays using multiple predictors and show that the extent to which diversion mechanisms affect operational outcomes at the facilities in the network depends upon the accuracy of the delay predictor used.
\noindent
\\
\\
We now briefly discuss the relevant literature. We first discuss diversion studies in the health operations literature. \cite{li2019review} summarized 137 articles addressing the ambulance offload delay problem and described how diversion reduced congestion and average wait times at a healthcare facility, and smoothens patient flow without increasing capacity \cite{nezamoddini2016modeling}. Centralized diversion policies are found to be preferable to decentralized policies, because decisions taken by one healthcare facility affect the operational outcomes of the other healthcare facilities involved in diversion \cite{baek2020centralized}. Therefore, sharing information regarding among healthcare facilities regarding their operational state becomes an important consideration during diversion. This is supported in the literature by studies that showed how diversion resulted in worse health outcomes due to lack of coordination within the diversion network, as diverted patients had to wait longer at the facility they were diverted to than at their facility of origin \cite{deo2011centralized}. Previous studies \cite{castillo2011collaborative, asamoah2008novel} have also proposed strategies to reduce or eliminate diversions such as increasing resource capacities at healthcare facilities. In \cite{ramirez2014optimal}, the authors presented a Markov decision process formulation for diverting ambulances in a two-facility problem, with the objective of minimizing the wait time of patients beyond a clinically important threshold duration. They assumed that the distribution of the time to start treatment at the other facility is known. In \cite{baek2020centralized}, the authors formulated the AD decision as an mixed integer linear program in terms of minimizing patient wait times across the entire diversion network. The formulation is solved at discrete time intervals on a rolling horizon basis and concluded that a formulation implementing a centralized policy outperformed other models in minimizing patient tardiness.
\noindent
\\
\\
It is thus evident that predicting delays on a real-time basis at all facilities in the diversion network can help develop a centralized diversion policy. We briefly discuss the real-time delay prediction literature in this context. Multiple studies have developed real-time delay predictors for arriving entities in different types of service systems and we refer readers to a relatively recent review \cite{ibrahim2018sharing} for a comprehensive account of the relevant literature. Three types of delay predictors have been proposed for predicting real time delays at service systems based on: (a) queue length \cite{whitt1999improving}, (b) delay history \cite{ibrahim2009real} and (c) machine learning \cite{baldwa2020combined}. With regard to M/G/1 queuing systems, while many delay predictors have been developed for this system, the existing delay predictors do not consider the limits or extreme quantiles of the service distributions in making their predictions, which we do in the delay predictor we develop in this study.
\noindent
\\
\\
Previous approaches have implemented diversion without providing personalized real-time delay estimates that patients might experience at each facility in the network. In this study, we predict delays on a real-time basis at both healthcare facilities we consider using system state (queue length, elapsed service time, etc.) and delay history-based delay predictors. We simulate the diversion mechanism using real-time delay predictions across both facilities. Our main contributions are: (i) to provide a framework for real-time delay prediction based diversion in a network of queueing systems, particularly in healthcare; (ii) to propose a simple and easy to implement delay predictor based on elapsed service time, queue length, and limits of the service time distribution, and (iii) to show how the accuracy of delay predictors is related to the extent to which operational outcomes become more equitable across the diversion network. We now briefly describe PHC operations, the health facility that we consider for diversion.

\section{Primary Health Centers}
In India, PHCs are the first point of contact with a formally trained doctor and cater to outpatients, and on a limited basis to inpatient, childbirth patients and those requiring antenatal care \cite{Guidelines}. We briefly describe the flow of patients through a PHC here. A detailed description of PHC operations, their simulation model development, including parameterization and outputs is provided in \cite{shoaib20}.
\noindent
\\
\\
A PHC typically contains one or two doctors (typically general physicians), a staff nurse serving inpatients and childbirth patients, another nurse assisting the doctors with outpatients, four to six beds for inpatients, and a labour room with one bed for childbirth patients. PHCs also house a clinical laboratory for conducting common laboratory tests and a pharmacy that also manages patient registration. The outpatient department (OPD) operates for eight hours a day whereas the inpatient and childbirth facilities operate on a 24X7 basis, with staff nurses working in shifts to manage these departments. Note that doctors are typically available only during outpatient hours but may attend inpatients and childbirth cases outside outpatient hours. Based on our visits to these facilities, this appears to occur very infrequently, and hence we assume in the model that doctors are available only during outpatient hours.
\noindent
\\
\\
Outpatients whose age is less than 30 years directly consult doctors upon arrival and if the doctor is busy, they join the outpatient queue. Patients whose age is greater than 30 years consult a nurse first before consulting the doctor. This nurse measures the patient’s vitals, including checking for hypertension and high blood glucose levels as part of a lifestyle and non-communicable disease prevention scheme. Once the patient has finished consulting with the doctor, a certain proportion of patients are sent to an in-house laboratory if tests are required. All patients exit via the pharmacy, where in addition to obtaining pharmaceuticals as required, patients also register their visit. Inpatients typically consist of those requiring admission and care for relatively simple conditions; more complex and/or life-threatening cases are referred to secondary or tertiary care facilities. Inpatient lengths of stay are limited by the facility to 24 hours. If inpatients arrive during OPD hours, they first consult with the doctor and they are then admitted to the inpatient ward where the staff nurse monitors their condition and provides care as required. Outside OPD hours, inpatients are attended to by the staff nurse directly. Upon arriving at the PHC, childbirth patients are also first attended to by doctors during OPD hours. The doctor initiates childbirth bed request for patients so that they can be sent to the labour room. Once the labour duration is finished, these patients are shifted to an inpatient bed for between 24-48 hours before they are discharged. Outside OPD hours, childbirth patients are attended to by the staff nurse while the rest of their patient flow remains the same. 
\noindent
\\
\\
Based on this patient flow and parameter estimates taken from \cite{shoaib20}, we simulate patient care operations at two PHCs. We program this discrete event simulation model of the operations of two PHCs in Python using the salabim package, on an Intel i7 64-bit Microsoft Windows OS with 16 GB memory. We present estimates of operational outcomes at both PHCs from the simulation in Table~\ref{tab1}. We only list outcomes relevant to this study – a full list is provided in \cite{shoaib20}.

\begin{table}[htbp]
  \centering
\begin{threeparttable}
  \caption{PHC Operational Outcomes}
    \begin{tabular}{|p{11.635em}|p{10.5em}|p{9.9em}|}
    \hline
    {Outcome Measures} & PHC 1 & PHC 2 \\
    \multicolumn{1}{|r|}{} & (4/1440/2880/1/1/1/6)* & (4/720/2880/1/1/1/6)* \\
    \hline
    Doctor occupancy & 0.627 (0.003) & 0.658 (0.004) \\
    \hline
    Staff nurse occupancy & 0.307 (0.005) & 0.447 (0.011) \\
    \hline
    Inpatient bed occupancy & 0.114 (0.004) & 0.209 (0.007) \\
    \hline
    Childbirth bed occupancy & 0.470 (0.019) & 0.923 (0.031) \\
    \hline
    \% of childbirth cases whose wait time exceeds 2 hours & 38.98 (2.77) & 88.19 (5.46) \\
    \hline
    \end{tabular}%
    \begin{tablenotes}[para,flushleft]
      \tiny
      \item $*$ Outpatient load/childbirth patient load /inpatient load /number of doctors/number of staff nurse per shift/labour bed/IPD beds
    \end{tablenotes}
  \label{tab1}%
  \end{threeparttable}
\end{table}%

\noindent
\\
We observe that a significant proportion of childbirth patients wait for more than a given threshold duration (assumed to be 2 hours) before being admitted to the labour room at the PHC they visit. It is this observation that motivated us to consider diversion for childbirth patients in this study. 

\section{Real Time Delay Prediction based Patient Diversion }

\begin{figure}[htb]
	{
		\centering
		\includegraphics[width=0.9\textwidth]{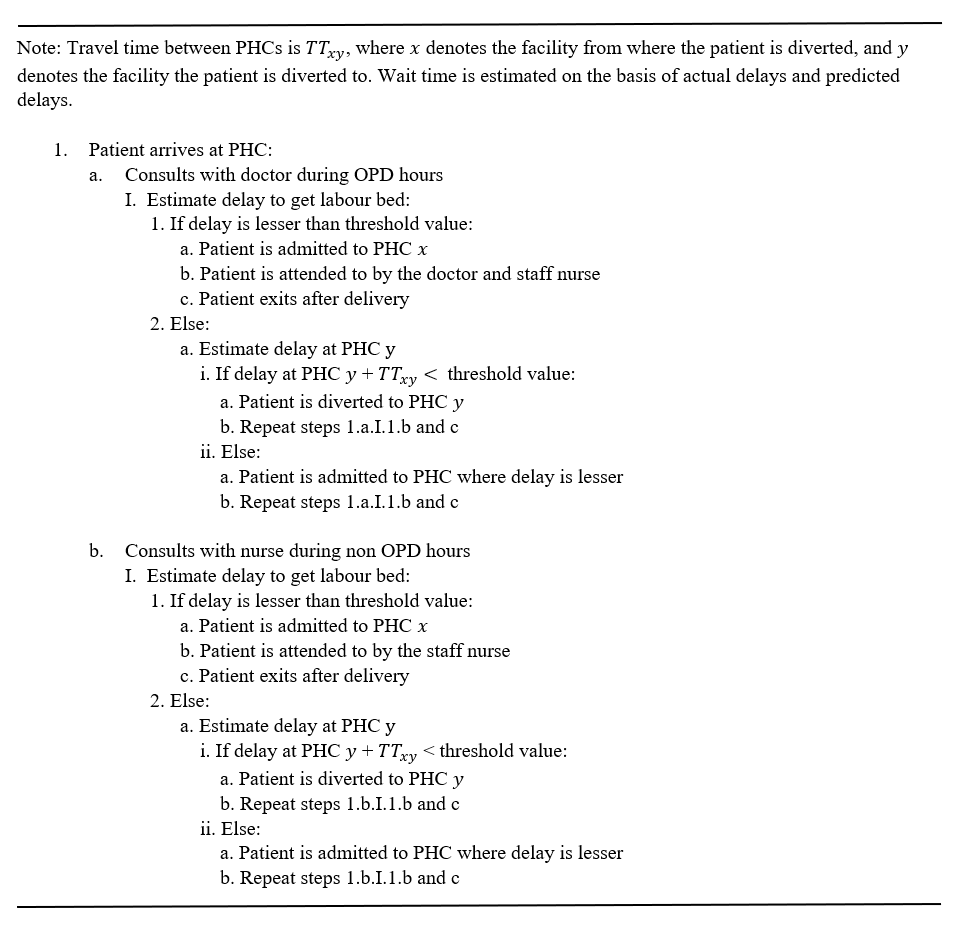}
		\caption{Diversion algorithm for childbirth patients.\label{fig1}}
	}
\end{figure}

In this section we present the diversion algorithm for childbirth patients and describe the delay predictors used for estimating the real-time delays of patients at both PHCs. In Figure~\ref{fig1}, we present the centralized diversion algorithm for childbirth patients. We also consider the travel time between PHCs during diversion and divert patients to the other PHC only when the predicted delay at the other PHC (estimated at the time the patient is expected to reach the other PHC) plus the patient’s travel time is lesser. Here we estimate expected delays of patients using system state variables such as queue length and elapsed service time.

\subsection{Delay Predictors}

We now describe the delay predictors used in our diversion model. We emphasize here that the delay prediction is made at the point in time when the patient arrives in the system - hence the term ‘real-time delay prediction’. Note that we do not use the average wait time as a delay predictor given that it is well established that average wait times are routinely outperformed by system state or delay history-based predictors for the purpose of real-time delay prediction \cite{ibrahim2018sharing}. The queueing system that we generate the delay prediction for is the labour room bed, which we approximate as an M/G/1 queueing system. The interarrival time distribution of the childbirth patients to the PHC is exponential with means for each PHC given in Table 1, and the time spent by the childbirth patient in the labour room bed (the “service time”) is uniformly distributed with parameters 360 minutes and 600 minutes. 

\noindent

1. Remaining service time-based delay predictor (predicted delay denoted as w$_{rst}$). 

\begin{equation*}
w_{rst} = L_{q} E[S] + Pr\left\{Server\ is\ busy\right\}E\left[remaining\ service\ time|server\ is\ busy\right]
\end{equation*}
\noindent
Here L$_q$ represents the length of the queue at the time the prediction is generated, and E[S] is the mean service time.
Upon simplifying, we get 
\begin{equation*}
w_{rst} = \frac{\ Pr\left\{Server\ is\ busy\right\} \ E\left[residual\ service\ time|server\ is\ busy\right]}{\left(1\ -\ \rho\right)}
\end{equation*}
\noindent
$\rho$ = fraction of time the server is busy (i.e., server utilization).
\noindent
\\
From \cite{gross2008fundamentals}, $ E\left[residual\ service\ time|server\ is\ busy\right] = \frac{E\left[S^2\right]}{2\ E\left[S\right]} = \frac{1\ +\ {C_B}^2}{2} E\left[S\right], \text{where}\ {C_B}^2\ =  \frac{var\left[S\right]}{E^2\left[S\right]} $.
\noindent
\\
Therefore, $w_{rst} = \left(\frac{1\ +\ {C_B}^2}{2}\right)\ \left(\frac{\rho}{1-\rho}\right)\ E\left[S\right] $.

\noindent

2.  Elapsed service time and service time distribution based (proposed). The predicted delay is denoted by w$_{est}$.

\begin{equation*}
w_{est} =  L_{q} E[S] + \max(( t_{avg} - t_{e},\ \min(t_{e} - t_{avg}, t_{max} - t_{e}))
\end{equation*}
\noindent
Here $t_{e}$ $= $ Elapsed service time of patient on labour room bed; $t_{min}$,\ t$_{max}$  = length of stay distribution limits on labour room bed ($t_{min}=$ 360 minutes, $t_{max}=$ 600 minutes); and $t_{avg}$  $=$   $\frac{t_{min} + t_{max}}{2}$.
\noindent

\section{Results}
In this section, we present simulation results using the delay predictors presented in Section 3.1 and compare operational outcomes for three cases: (a) no diversion, (b) actual delay-based diversion, and (c) delay prediction base diversion. We quantify accuracy of delay predictors using the mean absolute percentage error (MAPE), specified by:  $\frac{1}{N}$$\sum_{i=1}^{N}\left|\frac{A_i-P_i}{A_i}\right|$, where A$_{i}$ represents actual delay and P$_{i}$ represents predicted delays estimated using different delays predictors and \textit{N} is the number of patients in our sample.
\noindent
\\
\\
As described in section 2, we observe that a significant proportion of childbirth patients experience substantial wait time before getting admitted in the healthcare facility they visit. With the no diversion case, we observe that approximately 75.92\% of childbirth patients wait longer than two hours before getting admitted to the labour room and this proportion reduces significantly to 42.43\% with actual delay based patient diversion. We also estimate the extent to which differences in operational outcomes between PHCs change when diversion is implemented. We note that prior to diversion, PHC 1 (see Table~\ref{tab1}) had significantly lower resource utilization levels when compared to PHC 2. Diversion helps make the utilization levels across PHCs more equitable, and we show the extent to which this occurs for resource levels directly involved in provision of childbirth care depends upon the accuracy of delay predictors employed.
\noindent
\\
\\
Table~\ref{tab2} shows the results for childbirth patients when diversion is implemented using actual delay values (obtained from the simulation) and using the delay predictors w$_{rst}$ and w$_{est}$. The results are benchmarked against the case when no diversion is implemented. We report the percentage differences between resource utilizations involved directly in childbirth patient care, the average wait time for the labour bed and the proportion of patients whose wait time exceeds two hours ($\alpha$). It is evident that patient diversion improves outcomes in general – both $\alpha$ and the labour bed wait time decrease with diversion. We see that the differences in resource utilization decrease the most $($or in other words, become more equitable across both PHCs$)$ as the accuracy of the delay predictor increases, with the greatest change (most equitable) observed when actual delays are used (i.e., 100\% accuracy), and then decreases as the accuracy of the delay predictor decreases. The MAPEs of the delay predictors w$_{rst}$ and w$_{est}$ are 21.43\% and 11.34\% respectively. When we conducted a sensitivity analysis by increasing the arrival rate of childbirth patients, we observed MAPEs of 20.88\% and 12.61\% for w$_{rst}$ and w$_{est}$ respectively, indicating that similar trends are observed even as patient load increases.

% Table generated by Excel2LaTeX from sheet 'Sheet2'
\begin{table}[htbp]
  \centering
  \begin{threeparttable}

  \caption{Percentage difference in operational outcomes when patient diversion is implemented}
    \begin{tabular}{|l|c|c|c|c|c|p{5em}|}
    \hline
    \multicolumn{1}{|p{6.2em}|}{Diversion case} & $\Delta\rho_{doc}$   &  $\Delta\rho_{nurse}$     &  $\Delta\rho_{IPD}$     &  $\Delta\rho_{lb}$     & \multicolumn{1}{p{13.8em}|}{Labour bed wait time (minutes)} & $\alpha$ \\
    \hline
    \multicolumn{1}{|p{5.545em}|}{No diversion} & 4.73  & 31.3  & 45.73 & 49.07 & 88.31 & 75.92(6.43) \\
    \hline
    w$_{act}$    & 1.99  & 9.21  & 13.9  & 15.48 & 56.79 & 42.43(2.72) \\
    \hline
     w$_{rst}$ & 2.52  & 22.29 & 25.53 & 24.57 & 60.47 & 63.52(3.69) \\
    \hline
    w$_{est}$ & 2.2   & 13.97 & 19.74 & 20.1  & 48.37 & 54.59(1.24) \\
    \hline
    \end{tabular}%
    \begin{tablenotes}[para,flushleft]
      \tiny
      \item $\Delta\rho_{doc}$ = \textit{difference in doctor’s utilization}; $\Delta\rho_{nurse}$ $=$ \textit{difference in the staff nurse’s utilization}; $\Delta\rho_{IPD}$ $=$ \textit{difference in inpatient bed utilization}; $\Delta\rho_{lb}$ $=$ \textit{difference in labour bed utilization}; $\alpha$ $=$ \textit{proportion of childbirth patients with wait time $>$ 2 hours}
    \end{tablenotes}
  \label{tab2}%
    \end{threeparttable}
\end{table}%

\section{Conclusion}
In this work, we present a framework for implementing real-time delay prediction based patient diversion across two healthcare facilities in an Indian district. Our study shows that the extent to which operational outcomes become more equitable (that is, a facility with high congestion will see a decrease in congestion, and a facility with low utilization levels will see an increase in utilization) depend upon the accuracy of delay predictors.
\noindent
\\
\\
In addition to the fact that to the best of our knowledge, our work is the first study that considers real-time delay prediction as a basis for diversion in a health facility network, the approach that we propose may be used for diversion in general queueing systems as well – for example, in call centers. While there is a substantial body of research in the area of real-time delay prediction for queueing systems \cite{ibrahim2018sharing}, we have not come across a study that implements diversion based on delay predictions. Note that articles investigating the impact of delay prediction on reneging, balking and jockeying (for multi-server systems) are present in the literature \cite{ibrahim2018sharing}; however, these assume voluntary actions on the part of the entity waiting for service, and not a policy undertaken by the queue administration.  
\noindent
\\
\\
A key assumption in our work involves the presence of a centralized administration for the network of healthcare facilities that monitors and records system state information at all facilities in the network. Such an administrator would have to generate the delay predictions whenever a new patient arrives at any facility in the network, and then make the diversion decision based on the delay estimates. This would imply availability of the requisite information technology infrastructure to facilitate deployment of this diversion mechanism.
\noindent
\\
\\
Future avenues of research involve extension of this work to include the entire network of PHCs in a district, and potentially also secondary and tertiary levels of care. We also did not consider compliance of patients with the diversion decision, which can be investigated in further studies to determine its effect on overall network operational outcomes.

\bibliographystyle{IEEEtran}
\bibliography{mybibliography}
\end{document}